# BIG DATA TECHNOLOGY ACCELERATE GENOMICS PRECISION MEDICINE

HAO LI[1]

[1]Software Architect, Datacenter Health and Life Science, Intel Corporation, Shanghai, China
hao.h.li@intel.com

## ABSTRACT

*During genomics life science research, the data volume of whole genomics and life science algorithm is going bigger and bigger, which is calculated as TB, PB or EB etc. The key problem will be how to store and analyze the data with optimized way. This paper demonstrates how Intel Big Data Technology and Architecture help to facilitate and accelerate the genomics life science research in data store and utilization. Intel defines high performance GenomicsDB for variant call data query and Lustre filesystem with Hierarchal Storage Management for genomics data store. Based on these great technology, Intel defines genomics knowledge share and exchange architecture, which is landed and validated in BGI China and Shanghai Children Hospital with very positive feedback. And these big data technology can definitely be scaled to much more genomics life science partners in the world.*

## 1. INTRODUCTION

In genomics life science research, the data volume is going bigger and bigger, which is calculated as TB, PB or EB etc. For example, Genomics sequencing generates more than 1TB data per patient. During 2015, 1.65 million new patients in US generates more than 4EB data. In CNGB (China National Gene Bank), there is 500PB data volume deployed for now and it is estimated that volume will be increased by 5-10PB per year. In SCH (Shanghai Children Hospital) and Sjtu (Shanghai Jiao Tong University) Super Computing Center, there have hundreds of nodes for totally 30PB storage deployment. The key priority will be how to store and analyze data with optimized way and how to exchange and share the data with each other.

Intel defines and deploys scalable genomics knowledge share and exchange architecture, which is landed and validated in BGI China and Shanghai Children Hospital with very positive feedback. The solution stores data by using high density big data Lustre file system and leverages the cloud storage for hot and cold data efficient management. And solution architecture provides customized GenomicsDB engine for genomics variant call data search by position with very fast speed. Because genomics position is discrete, GenomicsDB also optimizes the sparse array storage by saving only the "useful" data. Intel also defines the genomics knowledge data sharing process and architecture for making the genomics knowledge be consolidated and utilized more efficiently.

## 2. INTEL BIG DATA ARCHITECTURE FOR LIFE SCIENCE

In real scenario, the research on genomics must work on big data mode. In this solution, it provides big data architecture (Figure 1). It is separated into 2 layer. One is application framework, which is likely interface to end user. It supports Genomics Knowledge App,

Genomics DB UI as well as Genomics Work Flow etc. The other is core framework and Linux kernel, which provides services to support request from application framework. The architecture supports big data level core framework. Such as TileDB and GenomicsDB Engine for big genomics variant data, Lustre file system with HSM (Hierarchal Storage Management) to leverage local and cloud storage for scaling to bigger data.

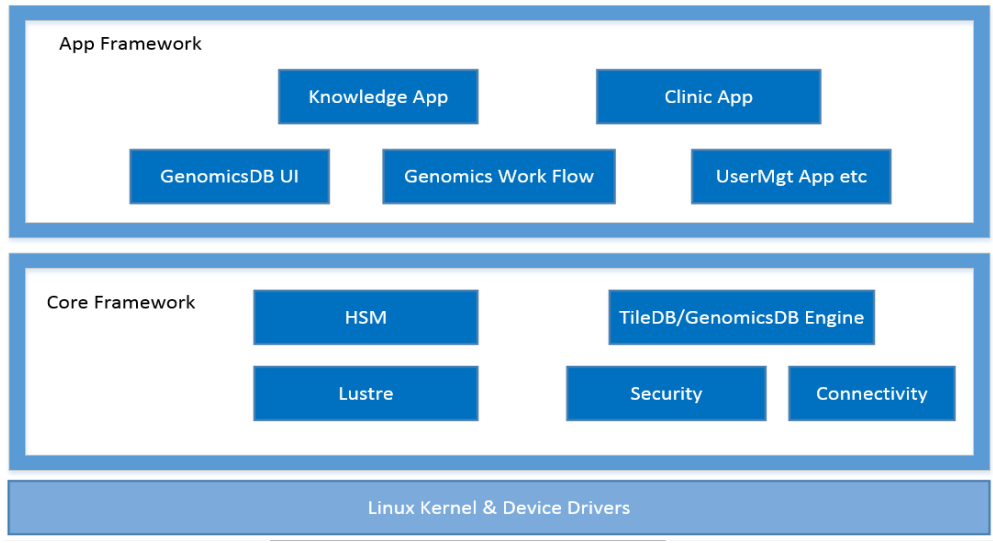

Figure 1. Big Data Architecture for Life Science

## 2.1. GENOMICS KNOWLEDGE SHARE MODEL

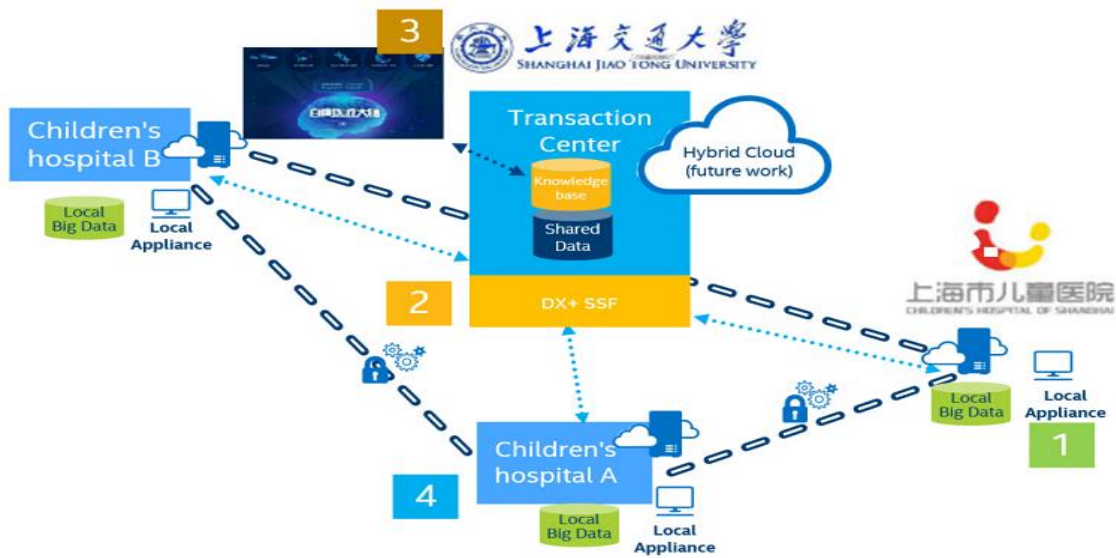

Figure 2. Genomics Knowledge Share Model

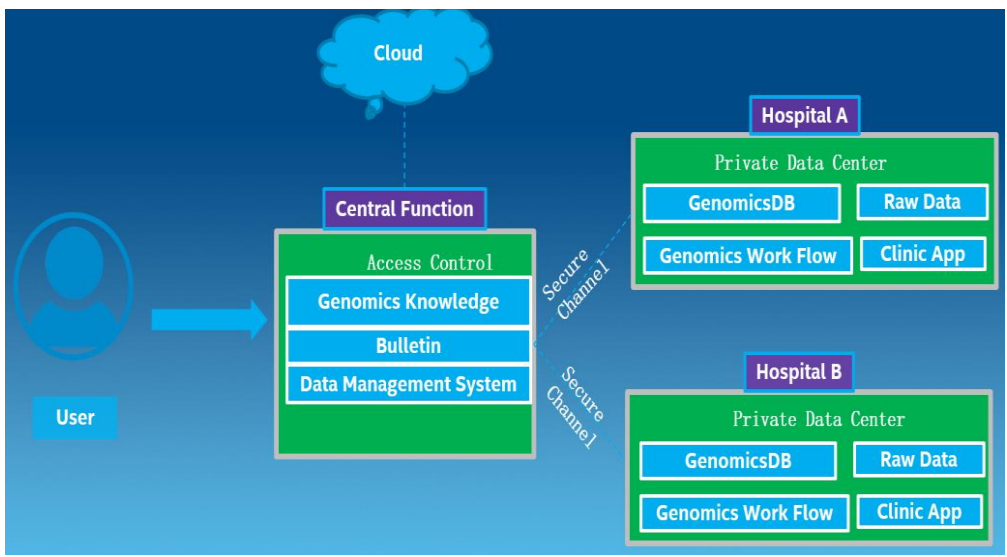

Figure 3. Genomics Knowledge Share Architecture

In Figure 2 and Figure 3, it shows key usage model and architecture of Genomics knowledge share. There is Central Function defined as data share agent. The purpose is for secure and consolidated data share and exchange. In real scenario, the genomics data is relate to privacy, hospitals can't share too much publicly. In this solution, hospitals keeps the raw data in local private data center, Central Function provides statistical and summarized genomics knowledge database share for query. For example, hospital A and B store raw genomics data with genomics work flow and create private GenomicsDB in their private data center. At the same time, hospital A and B can contribute statistical and representative variant call data to Central Function, data management system can create consolidated GenomicsDB knowledge center for share. Then end user can query consolidated GenomicsDB knowledge from both hospital A and B. And the data share is not for raw genomics data but statistical and GenomicsDB knowledge. Other than hospitals' private data center, Central Function can be deployed on public cloud with secured access control for data share.

**2.2. Visualized Genomics Work Flow**

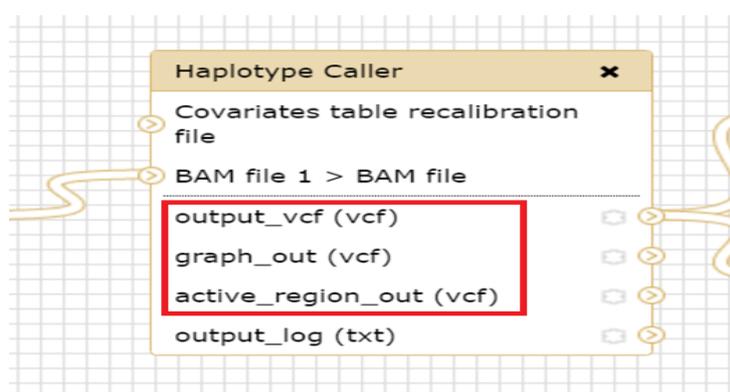

Figure 4. Visualized Genomics Work Flow

In Figure 4, it shows example of visualized genomics work flow, like data convert from fastq to BAM and then create final vcf (variant call format) data. In life science, some bio researchers don't have too much IT background. The visualized UI can help bio researcher much during customization of genomics data analysis and conversion. These raw data should be put in secure environment like hospital private data center and will be used to create GenomicsDB knowledge.

## 2.3. GenomicsDB and TileDB

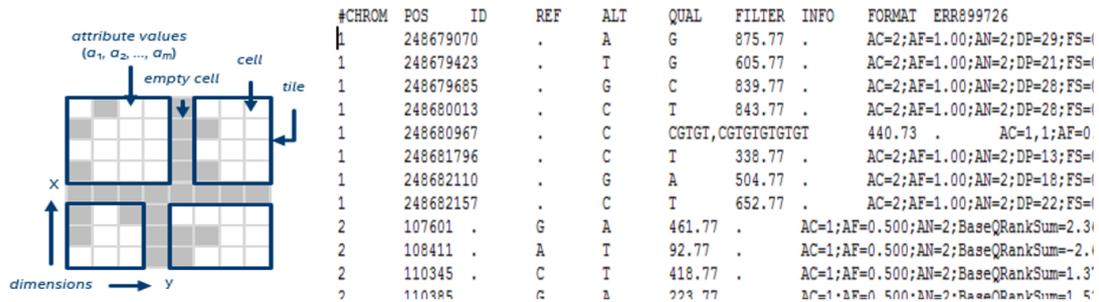

Figure 5. TileDB and GenomicsDB

In Figure 5, it shows TileDB work model. TileDB is a system for efficiently storing, querying and accessing sparse array data. It is optimized for sparse data and supports high performance linear algebra. For example, when storing data and querying cell, TileDB skips the empty cell to save much storage and query time. The GenomicsDB is instance of TileDB, which stores variant data in a 2D TileDB array. Each row corresponds to sample in a vcf and each column corresponds to a genomic position. Figure 5 also shows example of discrete genomics position data.

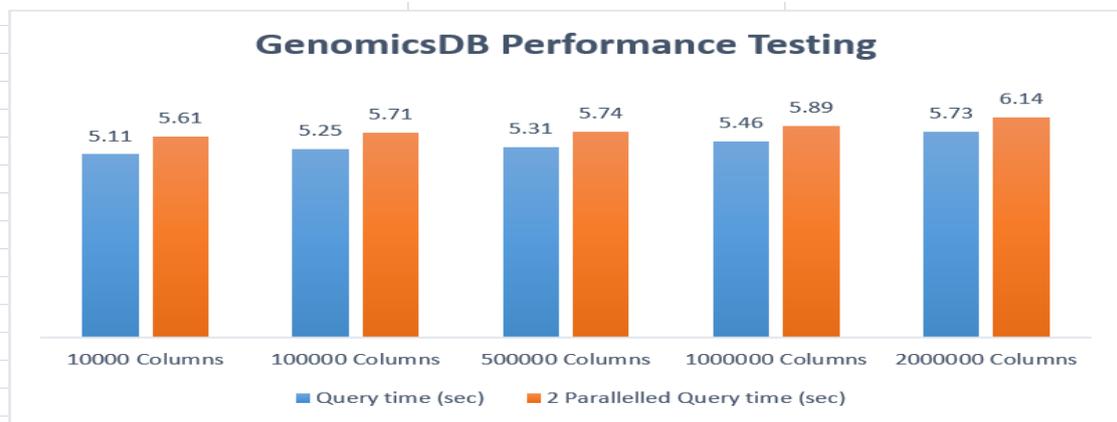

Figure 6. GenomicsDB Performance Report

Figure 6 shows real GenomicsDB testing report from shanghai children hospital with 11G sample vcf. It takes seconds time to response user and shows better performance when doing paralleled query and scaling to millions of variant data column range.

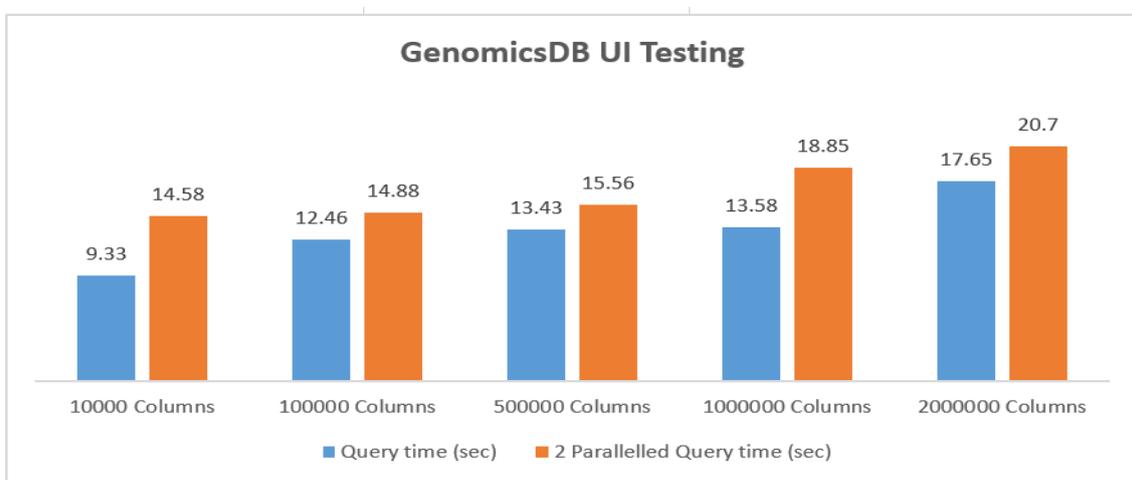

Figure 7. GenomicsDB UI

Figure 8 GenomicsDB UI Testing Report

Since some of bio researchers don't have much IT background, this solution provides very friendly web UI to support the query by end user. In Figure 7, it shows the convenient UI that has been landed in shanghai children hospital. In Figure 8, it shows the GenomicsDB UI testing report. Although it takes longer time to UI parse, render as well as annotation process etc, it still takes only seconds time to response user and shows great performance when scaling to bigger data with bigger column rage.

## 2.4. Lustre Big Data File System

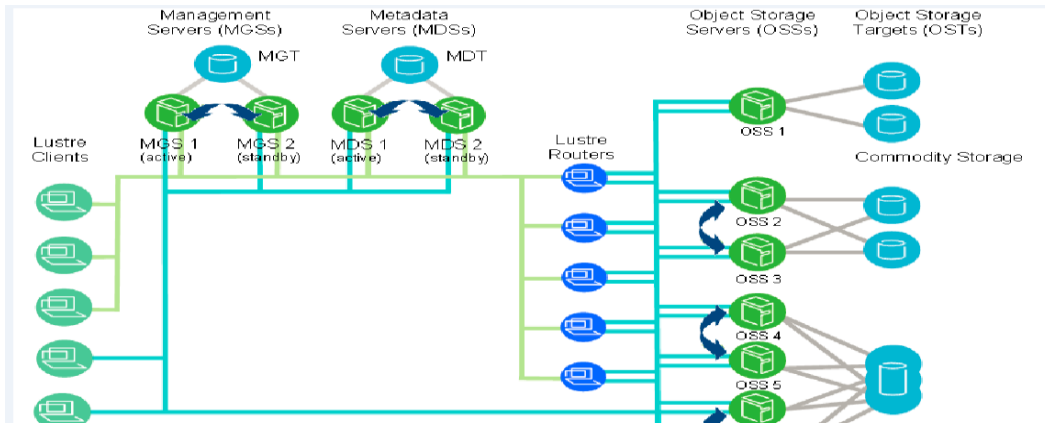

Figure 9. Lustre Architecture

Lustre file system is an open-source, parallel file system that supports many requirements of leadership class HPC simulation environments. Lustre is designed as high density file system, which utilizes physical storage as few as possible. It is easily scaled to bigger data in life science and supports standard file system behavior without additional protocol support. So it can be seamlessly integrated with legacy ecosystem. In Figure 9, it shows architecture of Lustre file system. The MGS (Management Servers) controls and manages entire Lustre file system. The MDS (Metadata Servers) manages the Meta data of multiple storages such as storage distribution, coordinators for data movers etc. The OSS (Object Storage Servers) manages the detailed storage devices. The Lustre Client is user interface, which acts as "normal" file system with POSIX interface. Once the client machine mounts Lustre file system, it can be treated as Lustre Client and Lustre details is transparent to end user.

## 2.5. Lustre HSM (Hierarchal Storage Management) Architecture

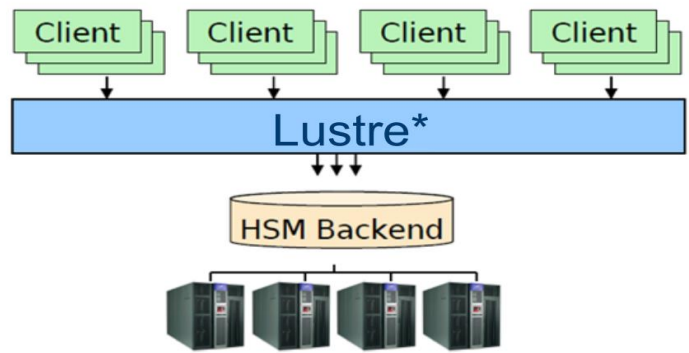

Figure 10. Lustre HSM Architecture

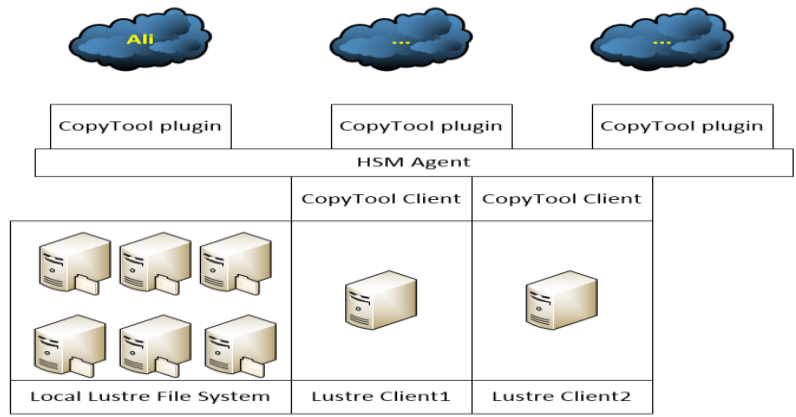

Figure 11. Scale by Plugin

Figure 10 shows Lustre HSM (Hierarchal Storage Management) Architecture, which supports bigger and bigger data. In this solution, hot data will be put on local Lustre and cold data will be put on cloud. HSM Agent controls the interface between different cloud services by plugin mode. And HSM Agent manages the file status system automatically, which makes correct action and status update be feasible and avoids incorrect file action that might causes to file missing. Furthermore, it supports parallelized upload and download with many Lustre clients. So as long as there is enough network bandwidth, the bandwidth of file data exchange between local Lustre and remote cloud is unlimited.

It can be scaled easily to new cloud services by creating plugin mode. For bio researchers, they don't need to get deep knowledge about how Lustre works. They just need to know the interface provided by HSM Agent and provide copy tool plugin then the tool can be scaled to more cloud services. See Figure 11.

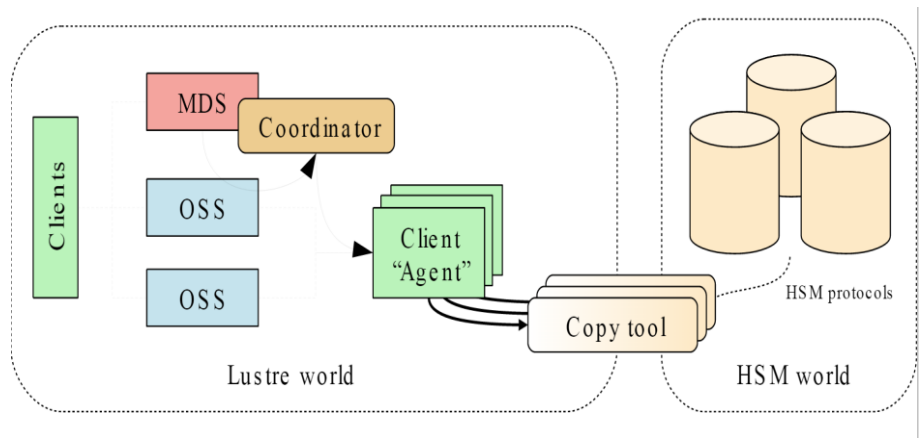

Figure 12. HSM Work Model

Figure 12 shows how Lustre HSM works. It utilizes Copytool as interface for data exchanging between local Lustre system and remote. The end users send request from copy tool client, then backend agent will work with Lustre internal coordinator to do the data exchange from/to remote cloud services by different HSM protocols.

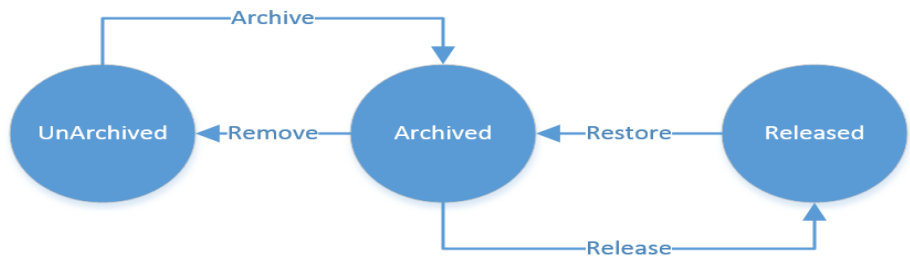

Figure 13. Standard Lustre Status Machine

Figure 13 shows standard Lustre file status machine

- Archive: Copy data from a Lustre file system file into the HSM solution.

- Release: Remove file data from the Lustre file system.

- Restore: Copy back data from the HSM solution into the corresponding Lustre file system file.

- Remove: Delete the copy of the data from the HSM solution.

Figure 14. Nature Language Interface

The solution supports nature language interface, which can make end users who don't have IT background be more comfortable. As it has been released as open source project, it can be easily to scale to more language support such as Chinese, Japanese, Korean, Spanish, Portuguese, and Russian etc. Figure 14 shows the example of nature language interface. For example, background part is used for checking runtime environment like runtime user, Lustre file system mount, HSM agent start etc. The scenario part is used for data exchange between local Lustre file system and remote cloud services.

## 3. CONCLUSIONS

As big data era of genomics life science industry is coming, previous traditional way can't fulfil the data growing request. By using Intel Big Data Architecture with GenomicsDB and Lustre, genomics data can be stored and utilized with optimized way. Central Function with genomics knowledge share will be scaled as future commercial standard. It can definitely facilitate life science research and accelerate the genomics precision medicine.

## ACKNOWLEDGEMENTS

Thank Carl Li, Guangjun Yu, Hui Lv, Jianlei Gu, Hong Sun, Ketan Paranjape, Paolo Narvaez, Karthik Gururaj, Kushal Datta, Danny Zhang, Julia Liang, Chang Yu, Ying Liu, Jian Li, Hua Ding, Zhiqi Tao, Hong Zhu, Ansheng Yang, Yan Li, Jinghai Zhang, Tijik Di, Hongchao Zhang and Robert Read for great help and support during solution design and development.

**Authors**

HAO LI

Hao is software architect in Intel Datacenter Health and Life Science Group. Hao joined Intel in 2005 and worked in many Intel internal and external open source product design and development such as Lustre HSM, GenomicsDB, Tizen, MeeGo etc. Hao received Master Degree of Computer Science from Fudan University in 2003.

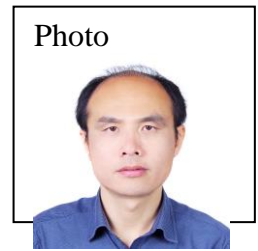

Photo